\documentclass[useAMS,usenatbib]{mn2e}

\include{epsf}
\usepackage{graphicx}
\usepackage{times}
\newcommand{\hi}{H\,{\sc i}}

\newcommand{\msol}{\mbox{${\rm M}_\odot$}}
\newcommand{\hubble}{\mbox{$\rm km\, s^{-1}\, Mpc^{-1}$}}
\newcommand{\kms}{\mbox{$\rm km\, s^{-1}$}}
\newcommand{\jykms}{\mbox{$\rm Jy\, km\, s^{-1}$}}
\newcommand{\mhi}{\mbox{$M_{\rm HI}$}}
\newcommand{\sint}{\mbox{$S_{\rm int}$}}

\newcommand{\ohi}{\mbox{$\Omega_{\rm HI}$}}
\newcommand{\mhis}{\mbox{$M^*_{\rm HI}$}}
\newcommand{\lmhis}{\mbox{$\log M^*_{\rm HI}$}}
\newcommand{\thetas}{\mbox{$\theta^*$}}

\newcommand{\vmax}{\mbox{$V_{\rm max}$}}

\title[HIPASS \hi\ mass function]{The HIPASS catalogue: $\Omega_{\rm HI}$ and environmental effects on the \hi\ mass function of galaxies}

\author[M.A. Zwaan et al.]{
M. A. Zwaan,$^{1}$\thanks{email:mzwaan@eso.org}
M. J. Meyer,$^{2}$
L. Staveley-Smith$^{3}$
and R. L. Webster$^{4}$\\
$^{1}$ European Southern Observatory, Karl-Schwarzschild-Str. 2, 85748 Garching
     b. M{\"u}nchen, Germany.\\
$^{2}$ Space Telescope Science Institute, 3700 San Martin Drive, Baltimore MD 21
     218, USA.\\
$^{3}$ Australia Telescope National Facility, CSIRO, P.O. Box 76, Epping, NSW 17
     10, Australia.\\
$^{4}$ School of Physics, University of Melbourne, VIC 3010, Australia.\\
}

\begin{document}

\date{Accepted 2005 February 9;
      Received 2005 February 1; in original form 2004 November 29}

\pagerange{\pageref{firstpage}--\pageref{lastpage}}
\pubyear{0000}

\maketitle

\label{firstpage}

\begin{abstract}
We use the catalogue of 4315 extragalactic \hi\ 21-cm emission line detections from
HIPASS to calculate the most accurate measurement of the \hi\ mass
function (HIMF) of galaxies to date.  The completeness of the HIPASS sample is 
well characterised, which enables an accurate calculation of space densities. The 
HIMF is fitted with a Schechter function  with parameters:
$\alpha=-1.37 \pm 0.03 \pm 0.05$, 
$\log (\mhis/\msol)=9.80 \pm 0.03 \pm 0.03\,h^{-2}_{75}$, and 
$\thetas=(6.0 \pm 0.8 \pm 0.6)\times 10^{-3} \,h^{3}_{75}\,{\rm Mpc}^{-3}$
(random and systematic uncertainties at 68\% CL),
in good agreement with calculations based on the HIPASS Bright Galaxy Catalogue, which is a complete, but smaller, sub-sample of galaxies.
The cosmological mass density of \hi\ in the local universe 
is found to be $\Omega_{\rm HI}=(3.5 \pm 0.4 \pm 0.4) \times 10^{-4} h_{75}^{-1}$.
This large homogeneous sample allows us to test  whether the shape of the HIMF depends on local galaxy density. We find 
tentative evidence for environmental effects  in the sense that
the HIMF becomes {\it steeper} toward {\it higher} density regions, ranging from $\alpha\approx-1.2$ in the lowest density environments to $\alpha\approx-1.5$ in the highest density environments probed by this blind \hi\ survey. This effect appears stronger when densities are measured on larger scales.

\end{abstract}
 

\section{Introduction}
The concept of the \hi\ mass function (HIMF) of galaxies was introduced 
by \citet{briggs1990} as a diagnostic tool for assessing the completeness
of optical galaxy catalogues against 21-cm \hi\ surveys. 
In that paper the then available luminosity functions and an adopted relation 
between gas richness $\mhi/L$ and luminosity $L$ were used to make the first preliminary HIMF calculations.
Since then much effort has been put into improving the calculations, 
mostly based on blind 21-cm \hi\ surveys because these -- by definition -- 
have  no bias against optically faint, gas-bearing galaxies 
\citep[][hereafter Z03]{zwaan1997,kraan1999,schneider1998,rosenberg2002,zwaan2003}.
The accuracy with which the HIMF can be evaluated
has improved significantly since Briggs' \citeyearpar{briggs1990} original calculation, 
but interestingly, the shape and normalisation of the field galaxy HIMF have  not changed
much. This is illustrative of the fact that the galaxy population seen in 
21-cm emission is essentially the same to that seen in the optical 
or IR wavelength, but obviously weighted toward the more gas rich galaxies,
and hence to the late morphological types. Two interesting conclusions can be drawn 
from the results of the 21-cm surveys:
 1) There does not seem to be a large population
of low surface brightness galaxies that is overlooked
in optical surveys \citep[cf.,][]{disney1976}. This conclusion is also reached by 
deep large area optical surveys such as the Sloan Digital Sky Survey \citep[SDSS,][]{blanton2001} and the Two degree Field Galaxy Redshift Survey  
\citep[2dFGRS][]{cross2002}, as well as by the Millennium Galaxy Survey \citep[MGC,][]{driver2005}.
2) There appear to be no -- or at least very few --  isolated self-gravitating \hi\ clouds without stars,
or ``dark galaxies''. These results are very important for our
understanding of the galaxy population of the local universe.

Once an accurate HIMF has been measured, it can be used to 
evaluate the total mass density of neutral hydrogen atoms in the
local Universe, $\Omega_{\rm HI}(z=0)$. At higher redshifts, the highest column density QSO absorption line systems are used to measure $\Omega_{\rm HI}$, and
these measurements have shown that the comoving neutral gas density 
at $z=2-5$ is approximately a tenth of a percent of the present critical density of the Universe
and about half of the mass density in stars at  the present epoch \citep[e.g.,][]{storrie-lombardi2000,peroux2003}.
For a good understanding of the evolution  of the Universe's \hi\ content, the measurement of a  local benchmark is obviously very important.

One aspect of the HIMF, its dependence on environment, has not been studied in detail
because samples of \hi-selected galaxies were not large enough to address this issue.
Recently, there has been much interest in the environmental effects on star formation rate \citep{lewis2002,balogh2004}, the luminosity function \citep{croton2004}, and luminosity, surface brightness and colour \citep{hogg2003}. Since the \hi\ properties of galaxies are intimately tied to the process of star formation,   gaining knowledge on the environmental effects
on \hi\ properties will help understanding the above mentioned results. Two previous studies discussed environmental effects on the HIMF: \citet{rosenberg2002} compared the field HIMF with that in a cluster region and \citet{springob2004} studied the HIMF of optically selected galaxies. In this {\em Letter\/} we will present for the first time a study of the environmental effects on the HIMF for a large sample of \hi-selected galaxies.

The measurement of an accurate HIMF
was one of the primary science drivers for the \hi\ Parkes All Sky Survey
(HIPASS). This survey, which blindly searched the whole southern sky
for 21-cm \hi\ line emission,  has now been completed and the catalogue
of 4315 extragalactic \hi\ signals has been released to the public 
\citep{meyer2004,zwaan2004}. A HIMF based on the 1000 HIPASS galaxies
with the highest \hi\ peak fluxes \citep{koribalski2004}, was published in Z03.  
This {\em Letter\/}  presents the 
state-of-the-art HIMF using the the full HIPASS catalogue. The velocity coverage
of HIPASS is $-1200$ to $12,700\,\kms$ and the typical noise level is 13 mJy beam$^{-1}$.
See \citet{meyer2004} for more details on the HIPASS catalogue. Throughout
this {\em Letter\/} we use $H_0=75 \, \hubble$ to calculate distance dependent quantities.

\section{Calculating the HIMF}
In Z03 we derived a bivariate stepwise maximum
likelihood (2DSWML) technique, which solves for the space density of
objects as a function of \hi\ mass (\mhi) and velocity width ($W$) simultaneously.
For bins $j$ in $\log \mhi$ and bins $k$ in $\log W$ the space density
$\theta_{jk}$ can be determined iteratively by evaluating the maximum likelihood 
solution derived by \citet{efstathiou1988}:
 \begin{equation}
 \theta_{jk}=
{n_{jk}}\left({\displaystyle\sum_{i=1}^{N_{\rm g}} H_{ijk}} 
 {/\Theta_i}\right)^{-1}, \label{theta.eq}
 \end{equation}
 where $n_{jk}$ is the number of galaxies in each bin,
$H_{ijk}$ is a function that causes the summation to only go over
over the galaxies that are within the volume accessible to sources in
bin $jk$. The denominator $\Theta_i$ is the integral over the bivariate \hi\ mass
function defined as
\begin{equation}
\Theta_i=\int_{M_{\rm HI}=0}^{\infty}\int_{W=0}^{\infty} 
  C(S',S'_{\rm int}) \theta(\mhi,W)\,d\mhi\,dW,
\end{equation}
where $C$ is a function that makes the integral go over the mass and
velocity range that source $i$ could have and still be detectable at
its distance $D_i$, and
\begin{eqnarray}
S'&=&S^i (\mhi/M^i_{\rm HI}) (W^i/W),\\
S'_{\rm int}&=&S_{\rm int}^i \mhi/M^i_{\rm HI}
\end{eqnarray}
are the effective peak flux and integrated flux of source $i$ if its
\hi\ mass would change from $M^i_{\rm HI}$ to \mhi\ and its velocity width
would change from $W^i$ to $W$.

In Z03 we specifically designed this method for a
peak-flux--limited sample. However, the equations given above are
generally applicable for samples constructed using different selection
criteria. The function $C$, which can be interpreted as a completeness
function, is defined by the selection criteria of the sample.  For
example, by setting $C=0$ for $S'$ smaller than the peak-flux--limit
and $C=1$ otherwise, the method is again equivalent to the
peak-flux--limited case in Z03.

It can be easily seen that the part between the parentheses in
equation~\ref{theta.eq} is essentially the effective volume available
to sources in bin $jk$.  The important modification that we make here
to the SWML method, is that instead of evaluating the summation per bin
$jk$, we calculate the effective volume for 
each galaxy individually by iteration:
\begin{equation} \label{veff.eq}
V^i_{\rm eff}=\displaystyle\sum_{j=1}^{N_{\rm g}} C(S'',S''_{\rm int}) 
 {/\Theta_i},
\end{equation}
where $S''$ and $S''_{\rm int}$ are defined as
\begin{eqnarray}
S''&=&S^j (M^i_{\rm HI}/M^j_{\rm HI})(W^j/W^i),\\
S''_{\rm int}&=&S_{\rm int}^j M^i_{\rm HI}/M^j_{\rm HI}.
\end{eqnarray}
Stable solutions for $V^i_{\rm eff}$ are found after $\sim 25$ iterations.
These values of $V^i_{\rm eff}$ are found {\em for each galaxy individually} and are
the maximum likelihood equivalents of \vmax\ for
the standard $\Sigma 1/\vmax$ method \citep{schmidt1968}. Summing the
values of $1/V^i_{\rm eff}$ in bins of \hi\ mass gives the original
2DSWML solution for the \hi\ mass function.  However, for maximum
likelihood estimators of the space density of objects, the
normalisation is lost and has to be determined afterwards. This is
done by calculating the mean galaxy density $\bar{n}$ via the minimum
variance estimator \citep{davis1982} and setting $\bar{n}$ equal
to the integral over the \hi\ mass function. To take into account the effects
of sample variance on galaxy density, the selection function is weighted 
by the inverse of the second moment of the two-point correlation function, which is set to $J_3=8000\, h_{75}^{-3}\, \rm Mpc^{3}$ \citep{meyer2005}. The selection function
$S$ is calculated similarly as in Z03.  Only the
distance range corresponding to $0.001 < S < 0.1$ is used in the
calculation of $\bar{n}$, to avoid adding noise from the farthest and
nearest regions of velocity space where the number of galaxies is low.
After this normalisation is applied to $V^i_{\rm eff}$, these values
represent true volumes. The method was tested extensively using synthetic samples,
similar to what is described in Z03.

\section{Results}
In principle, the 2DSWML technique can be applied to the full
HICAT sample, including the detections for which the expected completeness
is very low. However, including the detections for which $C<<1$ implies
large corrections to the expected space densities (see Eq.\ref{veff.eq}). To avoid this, 
we choose to
cut the sample at $C>0.5$. This reduces the total sample by only 7\% 
to 4010. For the completeness function $C$ we adopt the parametric 
approximation based on an error function calculated in \citet{zwaan2004}, and to account for the varying 
noise level across the sky, we adopt the completeness correction suggested
in that paper. We note that the latter correction has an insignificant effect on the 
HIMF. In \citet{zwaan2004} we found that
HICAT is complete at the 95\% (99\%) level at  a peak flux of 68 (84) mJy  and
at an integrated flux of 7.4 (9.4) $\rm Jy\, km\, s^{-1}$

\begin{figure}
\begin{center}
\includegraphics[width=8.3cm,trim=0.5cm 3.8cm 0.5cm 0.4cm]{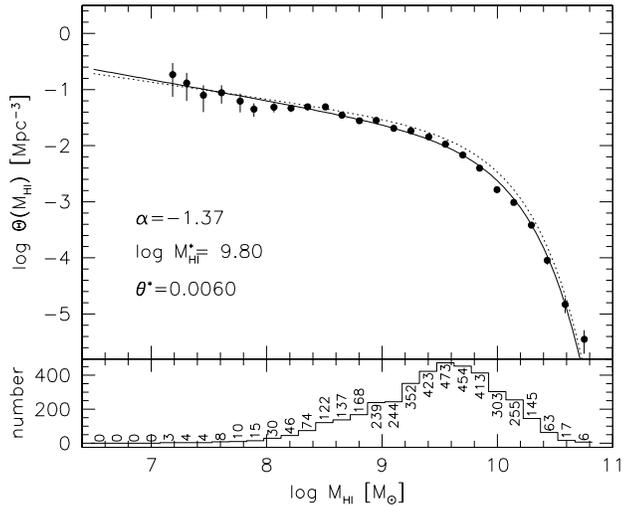}
\caption{{\em Top:\/} \hi\ mass function from HICAT. The solid line is a Schechter fit
to the points, the best-fit parameters are shown in the lower left corner; {\em Bottom:\/} Distribution of \hi\ masses used for the HIMF calculation. \label{hicat_himf.fig}}
\end{center}
\end{figure}

Figure~\ref{hicat_himf.fig} shows the result of the analysis. The top panel shows the HIMF, the 
bottom panel shows a histogram of the number of galaxies in each \hi\ mass bin.
The errors on the HIMF points are from Poisson statistics.
A Schechter function defined by 
\begin{equation}
\Theta(\mhi) d\mhi=\thetas \left(\frac{\mhi}{\mhis}\right)^\alpha \exp\left(-\frac{\mhi}{\mhis}\right) d\left(\frac{\mhi}{\mhis}\right),
\end{equation}
is fitted to the points and shown by a solid line. 
The parameters of this fit are:
$\alpha=-1.37 \pm 0.03$, 
$\log (\mhis/\msol)=9.80 \pm 0.03 \,h^{-2}_{75}$, and 
$\thetas=(6.0 \pm 0.8)\times 10^{-3} \,h^{3}_{75}\,{\rm Mpc}^{-3}$. 
The uncertainties on the Schechter parameters are determined by jackknife resampling \citep{lupton1993}, where the sky is divided into 24 equal area regions. 
This jackknife resampling procedure includes the uncertainties due to large-scale structure and
possible errors due to the varying noise level across the sky.
The dimensionless covariance between $\alpha$ and $\lmhis$ is $r(\alpha,\lmhis)=-0.53$, indicating the strong correlation between these two parameters, as is normally seen in galaxy luminosity and
mass function fits. We note that  \lmhis\ and \thetas\  are also highly covariant: $r(\lmhis,\thetas)=-0.83$, which illustrates that quoting values of \lmhis\ or \thetas\ individually is not very meaningful: the combination of these two parameters defines \ohi. 
For completeness, we note that  $r(\alpha,\thetas)=0.32$.

\begin{figure}
\begin{center}
\includegraphics[width=8.3cm,trim=0.5cm 0.7cm 0.5cm 0.4cm]{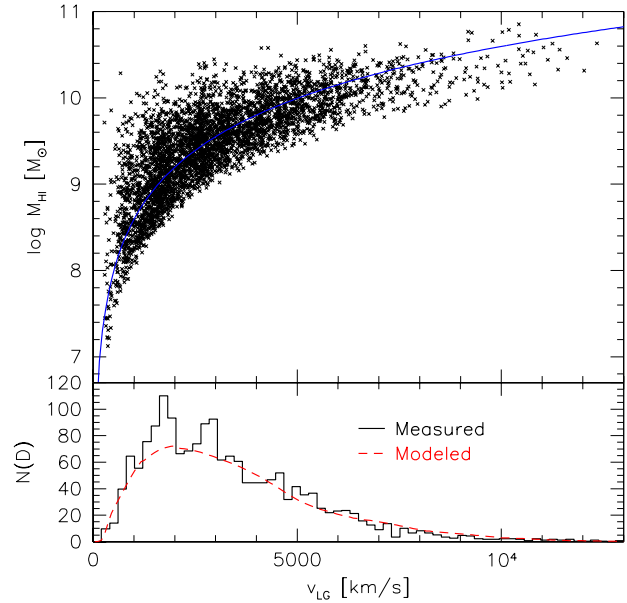}
\caption{{\em Top:\/} \hi\ masses of HICAT detections as a function of their redshift. Points above the solid line are integrated flux-limited at $\sint=9.4\,\jykms$ \citep{zwaan2004}. Only points that are used in the HIMF calculation ($C>0.5$) are shown. 
{\em Bottom:\/} Redshift distribution of HICAT galaxies. The histogram shows the measured distribution and the dashed line is predicted distribution based on the selection function calculated from the maximum likelihood method. \label{dist_hist.fig}}
\end{center}
\end{figure}

For comparison, the HIMF derived from the BGC (Z03) is plotted as a dashed curve.
The new HICAT HIMF is slightly steeper than the Z03 HIMF, for which we found 
$\alpha=-1.30 \pm 0.08$, but within the errors. The new measurement is also 
in reasonable agreement with previous HIMF measurements from blind 21-cm surveys,
which showed values of 
$\alpha=-1.2$ \citep{zwaan1997},
$\alpha=-1.52$ \citep{kilborn2000},
$\alpha=-1.51$ \citep{henning2000}, and
$\alpha=-1.53$ \citep{rosenberg2002}.
We refer the reader to Z03 for a complete
discussion of previous HIMF measurements from blind 21-cm surveys. 

In Figure~\ref{dist_hist.fig} we show the redshift distribution of the HICAT sample. A good
method to test the validity of the method used to calculate the HIMF is to
compare this distribution with the predicted distribution from the derived selection 
function $S$. This predicted distribution is shown by the dashed line, which appears to be an excellent approximation to the real HICAT selection.
Similarly to what we found in Z03, the number counts predicted from the selection function 
(0.174 deg$^{-2}$) is only $7\%$ lower than the measured number counts (0.186 deg$^{-2}$). This difference is due to the two over-densities seen at $\sim 1500$ and $\sim 2500~\kms$.

\subsection{Biases in the HIMF determination}
In Z03 we discussed in detail the various biases in the determination of
the HIMF based on the Bright Galaxy Catalogue. It was found that the selection bias
and \hi\ self-absorption have the strongest effects on the HIMF and $\Omega_{\rm HI}$.
For brevity, we choose not to repeat here all the tests of biases in the determination of the HIMF, but only briefly discuss the ones that may be particularly relevant to HICAT and refer the reader to Z03 for a detailed description of biases.

{\em Selection bias ---}
The selection bias is the effect that noise on detection spectra preferentially causes peak fluxes to be overestimated, which may result in an increased number of low-flux detections in the sample.
Similarly to Z03, we simulate this effect and find that due to the selection bias the global space density of objects is probably overestimated by $\approx 10\%$, and the slope of the HIMF becomes steeper by $\approx -0.05$. Together, this causes an overestimation of 
$\ohi$ of $\approx$15\%.

{\em \hi\ self absorption ---}
The effect of \hi\ self-absorption is very difficult to assess from the HIPASS data. We conservatively adopt the result from \citet{zwaan1997} that \hi\ self-absorption causes a
small decline in the measured space density and the knee of the HIMF, resulting in underestimation of the measured $\ohi$ of $\leq$15\%.

{\em Deviations from Hubble flow ---}
\citet*{masters2004} recently argued that calculating distances to galaxies by assuming
a pure Hubble flow instead taking into account the local velocity field, can lead to
errors in the slope of the HIMF. We test this by adopting the \citet{mould2000} distance model and recalculate the HIMF. The low-mass slope of this HIMF is  $\alpha=-1.39\pm 0.03$ 
not significantly different from our HIMF based on pure Hubble flow distances in the 
Local Group reference frame. \citet{masters2004} concluded that the Z03 HIMF slope
was probably underestimated and they suggest a slope of $\alpha=-1.4\pm0.1$, which is in excellent agreement with our new HICAT slope. The median distance of the HICAT sample
is larger than that of the Z03 sample, which implies that the fractional distance errors due to peculiar motions are smaller. This explains why taking into account the local peculiar velocity field does not have a significant effect on the HICAT HIMF slope.

{\em Lower distance limit --- }
To avoid confusion with Galactic high-velocity clouds (HVCs), we applied a 
lower limit of  $v_{\rm gsr}<300 \,\rm km \, s^{-1}$ to HICAT.  Obviously, this constraint
also excludes very nearby, low mass galaxies from the present analysis. 
Our previous HIMF 
determination based on the Bright Galaxy Catalogue (Z03) had a measurement  at $\mhi=10^{6.8}\msol$, but the lowest \hi\ mass point for HICAT is at $\mhi=10^{7.2}\msol$.
In Z03 we argued that for  peak-flux--limited samples, imposing a lower distance limit can cause a slight drop of the HIMF at very low \hi\ masses. For HICAT we find that this effect is not measurable
because the sample is not strictly peak-flux--limited.

\section{Environmental effects on the HIMF}
\citet{rosenberg2002} speculated on the effects of environment on the slope
of the HIMF. They concluded that  the HIMF in the immediate region around the Virgo cluster was $\alpha \approx -1.2$, flatter than in the field region where they found $\alpha \approx -1.5$. Our sample of HICAT galaxies is sufficiently large to 
test the effect of local density on the shape of the HIMF, by subdividing the catalogue
into samples drawn from different local densities. 

 We choose to parameterise local density by calculating the distance $r_n$ to the $n$-th nearest neighbour for all HICAT galaxies. Densities are calculated as $\rho_n=n/[(4/3)\pi r_n^3]$, and normalised to the median density to obtain $\rho_n/<\rho_n>$.
We divide the HICAT sample into five logarithmically spaced subsamples with different mean densities.
Since HICAT is increasingly sparsely sampled at larger redshift, 
nearest neighbour measurements increase rapidly as a function of redshift.
Therefore, different subsamples could effectively be drawn from different redshift ranges. This, in turn, implies preferential selection of certain \hi\ mass ranges for each subsample (see Fig.~\ref{dist_hist.fig}), which could bias the HIMF calculation. 
We correct for this effect by first fitting a power-law to the conditional probability distribution of nearest neighbour distance as a function of redshift. Next we subtract this power-law fit from
the nearest neighbour distances to obtain an unbiased nearest neighbour measurement of
which, however, the absolute calibration is now lost.
We test that our corrected nearest neighbour measurements sample all distances equally by calculating for each subsample the mean and the dispersion of the redshifts. We confirm that there are no significant differences in the redshift distributions of the subsamples.  Finally, to prevent including galaxies from the most sparsely sampled redshift range, 
we limit our calculations to galaxies with $v_{\rm gsr}<5000\,\rm km\,s^{-1}$.

\begin{figure}
\begin{center}
\includegraphics[width=8.3cm,trim=0cm 0cm 0cm 0cm]{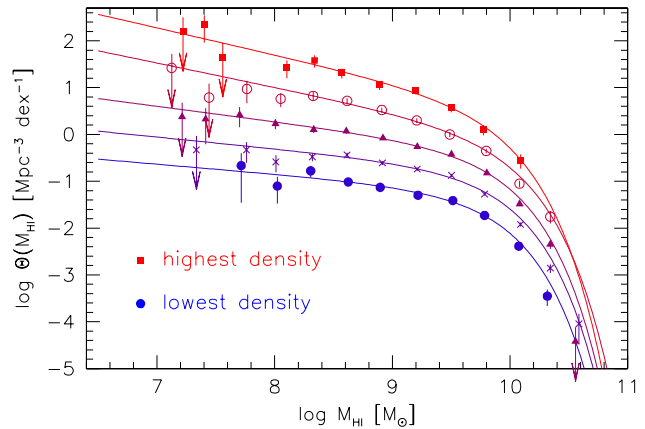}
\caption{The \hi\ mass function as a function of local galaxy density. The HICAT sample has been divided into five logarithmically spaced subsamples according to the density out to the fifth nearest neighbour. The HIMF for each subsample is shown with the corresponding best bit Schechter function. The HIMFs are offset from each other vertically by 0.5 dex. 
\label{himfden.fig}}
\end{center}
\end{figure}

\begin{figure}
\begin{center}
\includegraphics[width=8.3cm,trim=0cm 0cm 0cm 0cm]{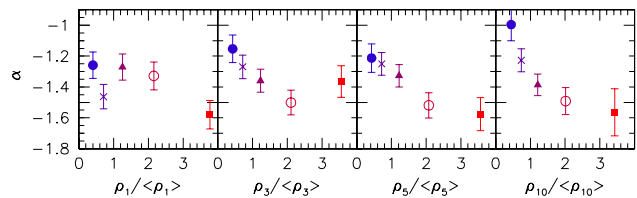}
\caption{The low-mass slope $\alpha$ of the Schechter function fits to the \hi\ mass function as a function of local galaxy density. The different panels show the results based on the first, third, fifth, and tenth nearest neighbour, from left to right.
\label{alphaden.fig}}
\end{center}
\end{figure}

The HIMF as a function of local density $\rho_5/<\rho_5>$ is shown in Fig.~\ref{himfden.fig}. We find 
evidence for a weak dependence of the HIMF  slope on density environment, in the sense that
the low-mass end of the HIMF becomes {\em steeper in higher density regions}.
There is  no corresponding correlation between the density and the characteristic mass \mhis. In addition to this analysis based on the fifth nearest neighbour, we also calculated densities on smaller and larger scales, out to the first, third, and  tenth nearest neighbour, respectively. The results based on these measurements are 
shown in Figure~\ref{alphaden.fig}, where the Schechter low-mass slope $\alpha$ is plotted as a function of density. Interestingly, the effect of steeper HIMFs toward higher density regions appears to become stronger toward larger scales, indicating that the environmental effects on the \hi\ properties of galaxies are not restricted to short distance effects such as galaxy interactions, but
also operate on larger scales.

Our observed trend between the slope of HIMF low-mass end with density is in qualitative
agreement  with \citet{croton2004}, who use the 2dF galaxy redshift survey
to show that the slope of  the luminosity function is weakly dependent on environment.
Our results cannot be quantitatively  compared to those of \citet{croton2004} because we use a different method to parametrise local density. Furthermore, our density measurements are based on 
\hi-selected galaxies only, while \citet{croton2004} use optically selected galaxies. 
It should also be kept in mind
that the dynamic range in densities probed by HIPASS is smaller than that
probed by optical surveys. For example, \citet{waugh2002} showed that 
cluster centres that are readily apparent in the counts of optical galaxies are only detected as over-densities of a few times the density of \hi-detected galaxies in the field.
At first sight, our findings seem to be in conflict with the recent results of 
Springob et al. (2004),
who, based on optically selected galaxies, claim a flattening of the HIMF toward higher density regions. However,  the statistical significance of their result is low and the HIMFs 
from the highest and lowest density subsamples are consistent at the $1\sigma$ level. 

Interestingly, recent 2dF and SDSS results show an anti-correlation between star formation rates based on the equivalent width of the  H$\alpha$ emission line and galaxy density \citep{lewis2002,gomez2003,balogh2004}, out to large distances from cluster cores. \citet{kauffmann2004} and \citet{tanaka2004} also find that the specific star formation rate ${\rm SFR}/M_*$, is strongly dependent on environment, but furthermore conclude that this relation is most pronounced for the lower mass galaxies. 
This implies that the comparative rate at which lower mass galaxies consume their innate \hi\ gas with respect to higher mass galaxies is greater in low density environments than high density environments.  Hence, this effect leads to a flattening of the HIMF in low density regions.
%
%
Of course, this is a simplified view of galaxy evolution because it is well established that
gas accretion plays an important role in the total neutral gas budget of galaxies \citep[e.g.,][]{matteucci1989}, but it can go a long way in explaining our observed correlation between HIMF slope and density. 

An alternative explanation can be found in the halo-occupation model in which galaxy properties depend solely on dark matter halo mass, which is predicted by the cold dark matter (CDM) theory.  \citet{mo2004} predict that in this model the shape of the luminosity function of late type galaxies is independent of local density because these galaxies live in low-mass halos. Early type galaxies mainly exist in high-mass halos, which are associated with higher density regions. It is known that the \hi\ masses of early type galaxies are typically low, and their HIMF is steeper than that of late type galaxies \citep{zwaan2003}. Therefore, a relatively higher contribution of lower \hi\ masses (and hence a steeper HIMF) in higher density regions is a direct prediction of the CDM model.


\section{Conclusions}
In Z03 we presented the \hi\ mass function of galaxies based on 
the  1000 HIPASS galaxies with the highest peak fluxes. That paper gives a 
detailed description of the HIMF calculation, the possible effects that may 
bias the calculation, and discussed the results. This {\em Letter\/}
presents an updated HIMF measurement from the full HIPASS sample.
We find
$\alpha=-1.37 \pm 0.03 \pm 0.05$, 
$\log (\mhis/\msol)=9.80 \pm 0.03 \pm 0.03\,h^{-2}_{75}$, and 
$\thetas=(6.0 \pm 0.8 \pm 0.6)\times 10^{-3} \,h^{3}_{75}\,{\rm Mpc}^{-3}$
(random and systematic uncertainties at 68\% confidence),
in good agreement with the Z03 results. The slope of the HIMF is comparable to that
of the optical luminosity function from the SDSS measured down to low absolute magnitudes 
of $M_r \sim -13$ \citep{blanton2004}.

We find tentative evidence for a steepening of the low-mass end of the HIMF as a function of local galaxy density. This effect becomes stronger if local densities are measured over larger scales, which indicates that environmental effects on the \hi\ properties of galaxies are not restricted to short distance effects such as galaxy interactions. 

HIPASS is a relatively shallow survey compared to some of the small area deep surveys.
However, 
because of the large sky coverage
the number of detections is very large and  the HIMF can be measured
with high accuracy. HIPASS has particularly  good statistics at  the high mass end
of the HIMF, which dominates the mass density of neutral hydrogen.
Therefore, \ohi\ is measured with high accuracy: $\Omega_{\rm HI}=(3.5 \pm 0.4 \pm 0.4) \times 10^{-4} h_{75}^{-1}$.

The faint-end tail of the HIMF is measured well down to \hi\ masses of $10^{7.2}\msol$.
No other survey has been able to constrain the HIMF better at this mass limit. The space density of objects much below $M_{\rm HI}=10^7\msol$ is still not
accurately known. A number of surveys have been able to put upper limits
on the space density of these low \hi\ mass systems in higher density regions
\citep[e.g.,][]{deblok2002,zwaan2001}, but
large scale `blind' surveys 
with existing technology are probably not able 
to make significantly better measurements in this region of parameter space.
Future large radio telescopes are required to measure reliably the space density of objects
much below $\mhi<10^7~\msol$.

\small
\bibliographystyle{mn2e}
\bibliography{mn-jour,all}

\label{lastpage}

\end{document}